\documentclass[sigconf,table]{acmart}

\usepackage{color, colortbl}
\usepackage{url}
\usepackage{multirow}
\usepackage{hyperref}

\usepackage{graphicx}
\usepackage{hyperref}

\usepackage{booktabs} %
\usepackage{csquotes}
\usepackage[capitalise,nameinlink]{cleveref}
\usepackage{paralist}
\usepackage[gen]{eurosym}

\newcommand{\eg}{e.\,g.,\ }

\newcommand{\lastAccessed}{, accessed: \today}

\definecolor{Gray}{gray}{0.925} %
\definecolor{Preprint}{rgb}{.63,.79,.95}

\acmConference[]{}
\acmBooktitle{}

\renewcommand\footnotetextcopyrightpermission[1]{} %
\usepackage{color, colortbl}
\usepackage[pscoord]{eso-pic}
\newcommand{\preprintBanner}{
\AddToShipoutPictureFG*{\put(\LenToUnit{0.5\paperwidth},\LenToUnit{0.95\paperheight}){\makebox[0pt][c]{
\renewcommand{\arraystretch}{1.5}
\setlength{\tabcolsep}{18pt}
\rowcolors{1}{Preprint}{Gray}
\begin{tabular}{|p{0.55\paperwidth}|} \hline
\textbf{Preprint from \texttt{\href{https://ostendorff.org/pub/}{https://ostendorff.org/pub/}}} \\ \hline
\footnotesize
M. Ostendorff, T. Blume, S. Ostendorff, ``Towards an Open Platform for Legal Information'' in \textit{Proceedings of the ACM/IEEE Joint Conference on Digital Libraries (JCDL)}, 2020. \\ \hline
\end{tabular}}}}
}

\definecolor{Gray}{gray}{0.925}

\begin{document}

\fancyhead{}

\title{Towards an Open Platform for Legal Information}

\author{Malte Ostendorff}
\affiliation{
   \institution{Open Justice e.V}
}
 \email{mo@openlegaldata.io}

\author{Till Blume}
 \affiliation{
   \institution{Open Justice e.V.}
}
 \email{tb@openlegaldata.io}

\author{Saskia Ostendorff}
 \affiliation{
  \institution{Open Justice e.V.}
}
\email{so@openlegaldata.io}

\renewcommand{\shortauthors}{Ostendorff et al.}

\begin{abstract}
Recent advances in the area of legal information systems have led to a variety of applications that promise support in processing and accessing legal documents. 
Unfortunately, these applications have various limitations, \eg regarding scope or extensibility.
Furthermore, we do not observe a trend towards open access in digital libraries in the legal domain as we observe in other domains, \eg economics of computer science.
To improve open access in the legal domain, we present our approach for an open source platform to transparently process and access Legal Open Data.
This enables the sustainable development of legal applications by offering a single technology stack.
Moreover, the approach facilitates the development and deployment of new technologies.
As proof of concept, we implemented six technologies and generated metadata for more than 250,000 German laws and court decisions.
Thus, we can provide users of our platform not only access to legal documents, but also the contained information.
\end{abstract}



\maketitle

\preprintBanner

\section{Introduction}
\label{sec:intro}

The importance of automatically processing legal documents is rising.
Recent advances in research offer a portfolio of technologies to process legal documents, \eg extracting, aggregating, and linking information from text.
Hence, mostly commercial tools and platforms have emerged that promise support in processing and accessing legal documents. 
Unfortunately, there are various limitations when using these tools.
For instance, they are country-specific, lack transparency and extensibility (closed source), or do not provide access to the raw data.
These criteria are essential for legal data analysis~\cite{Fleckner2018}, and the development of innovative technologies, \eg visual query interfaces\footnote{\url{https://www.vizlaw.de}\lastAccessed}.
Data analysis and visualization are of great benefit when interpreting the information, \eg to investigate the mutual dependencies between statutes and the temporal evolution of law.
In our opinion, democratizing the access to these tools and providing the data is fundamental when one is interested in facilitating access to justice and innovation in the legal domain.
A key element to achieve these goals is open data, that can reduce integration costs, improve transparency, and harness the innovation of others~\cite{Shadbolt2011}.

In this paper, we present our approach for an open source platform to transparently process Legal Open Data by flexibly combining state-of-the-art technologies.
Our approach enables the sustainable development of legal data processing tools by offering a single technology stack.
The platform empowers others to quickly develop and deploy new technologies.
As proof of concept, we implemented six technologies in an open processing pipeline, processed more than 250,000 laws and court decisions, and made them available on our Open Legal Data Platform.
Our source code and our generated data is publicly available \footnote{\url{http://www.openlegaldata.io/}\lastAccessed}.

Below, we briefly discuss representative related projects and platforms.
Subsequently, we present our approach as well as the implemented technologies.

\section{Related Work}
\label{sec:related-work}
CourtListener\footnote{\url{http://www.courtlistner.com}\lastAccessed} is a service for the United States, which is developed by the non-profit Free Law Project. %
CourtListener's goal is \enquote{to provide free, public, and permanent access to primary legal materials on the Internet for educational, charitable, and scientific purposes to the benefit of the general public and the public interest}~\cite{Lissner2010}.
CourtListener seeks to collect and freely distribute historical and current United States court opinions on state and federal level.
However, other international jurisdictions are not in the scope of the project.  
Similarly, the Caselaw Access Project\footnote{\url{https://case.law}\lastAccessed} by Harvard Law Library aims to make all published U.S. court decisions freely available. 
The Finnish government developed the web service Finlex \footnote{\url{http://www.finlex.fi}\lastAccessed}, which provide laws and related legal documents as XML documents.
In 2014, Frosterus et al.~\cite{Frosterus2014} improved Finlex in several ways, \eg by transposing the XML documents to RDF documents following the Linked Open Data principles.
They demonstrate the usefulness of Linked Open Data for content producers, application developers, and data analysts.
OpenLaws\footnote{\url{http://www.openlaws.eu}\lastAccessed} is an open access platform for European legal information~\cite{Lampoltshammer2016}.
OpenLaws is built on top of open source software, but it does not provide access to the data.
In summary, there are various projects scattered across the world that collect and publish legal documents.
However, there is no single project that is open source, makes data openly accessible, and is not focused on a single country only.

\section{The Open Legal Data Platform}
\label{sec:approach}
\begin{figure}[!bt]
\centering
\includegraphics[page=1,width=.5\textwidth,trim=1.4cm 1.2cm 1cm 2cm, clip]{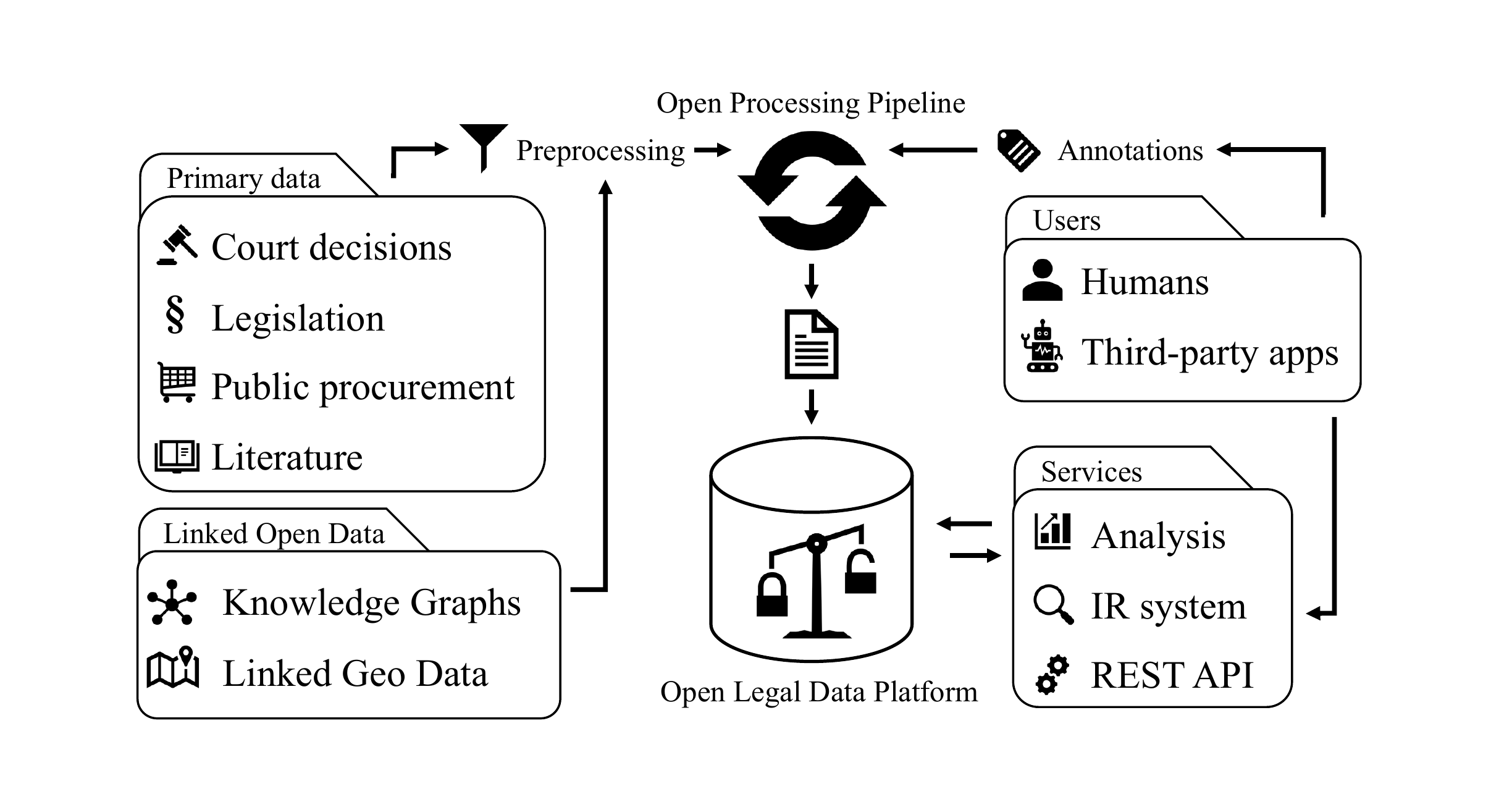}
\caption{\label{fig:high-level}The Open Legal Data Platform vision: Harvesting legal documents from official sources, preprocessing of raw documents and enriching the documents with data from additional sources. The platform content is accessible through an API to facilitate analysis, information retrieval, and third-party apps.}
\end{figure}
Our approach to develop a single legal technology platform is illustrated in \cref{fig:high-level}.
We provide the basic technology stack that legal engineers can build upon to develop new technologies.
The developed technologies can be flexibly combined to provide country specific-platforms, \eg for Germany.
Developed technologies can be included in the global technology stack to make them accessible to developers around the world.
Moreover, legal tech developers benefit from the tools and the data provided by the platforms via our REST API.
Finally, researchers benefit from easy access and the latest technologies integrated into a single platform ecosystem to conduct analyses.

Although there are country-specific differences in the legal systems, we firmly believe that the development of a legal technology platform can be tackled with a single technology stack.
Akoma-Ntoso~\cite{Palmirani2011} is one prominent example for an XML schema that aims to standardize legal documents on an international level. 
By integrating standards, like Akoma-Ntoso, the platform comparability for different countries can be facilitated.
For any Open Legal Data platform, we need to access primary and secondary data sources (\cref{fig:high-level} left), process the data to generate additional information (\cref{fig:high-level} center), and provide services to users (\cref{fig:high-level} right).

For Open Legal Data, we access primary source like government services, and secondary sources liked Linked Open Data.
The preprocessing system is designed to flexibly handle different types of documents, such as legislation or literature.
The processing pipeline allows to enrich documents with additional information, \eg automatically extracted references or manually created text annotations from domain experts.
Finally, the data is made available to the public via information retrieval systems or REST APIs.
As a foundation, we use the Django framework \footnote{\url{http://www.djangoproject.com}\lastAccessed}.
The user interfaces can be translated into different languages and adapted to specific information needs using Django's template system.
Django's \enquote{app system} enables easy integration of new modules and the re-use of existing apps. 

In the following, we describe in more detail technologies to access, process and provide legal documents.
All technologies are integrated or are currently being integrated into our Open Legal Data technology stack. 

\subsection{Primary data sources \& Linked Open Data}
\label{sec:german-law}
Finding and harvesting legal information is a challenging task due to several reasons. 
Accessing data directly from courts is time-consuming and expensive.
Accessing data from sources on the Web induces quality issues.
In the following, we present three alternative sources of legal information and our approach to include them.

\paragraph{Courts and Governmental data.}
We collaborate with courts to obtain decisions directly.
Accessing data directly from courts has the highest level of trustworthiness.
However, it is very time consuming since courts do not always make decisions publicly accessible.
Furthermore, decisions obtained directly from courts are rarely in machine-readable formats and are not free of charge.
Thus, information needs to be extracted from, \eg from purchased PDF files.

\paragraph{Crawling trusted websites.}
Crawling trusted websites can significantly improve the amount of data.
In Germany, there exists a small set of trusted websites.
The German Federal Ministry of Justice and Consumer Protection (BMJV) operates websites with the latest version of federal legislation\footnote{\url{http://gesetze-im-internet.de}\lastAccessed} and decisions from federal courts\footnote{\url{http://rechtsprechung-im-internet.de}\lastAccessed}. 
In Germany, state-level legislation and decisions are not available on a central web service.
Each state needs to be handled separately.
On European level, the service EUR-LEX\footnote{\url{http://www.eur-lex.europa.eu}\lastAccessed} is the main data source.
Additionally, we crawl legal blogs which have been shown to provide information for legal opinion mining~\cite{Conrad2007}.
Having different data sources requires the harmonization of the harvested data to avoid duplicates.
To de-duplicate court decisions, we use the European Case Law Identifier~\cite{VanOpijnen2011}.

\paragraph{Linked Open Data.}
According to the Open Data Monitor \footnote{\url{http://www.opendatamonitor.eu/}\lastAccessed}, 45\% of all Open Data is currently provided in (semi-)structured and thus machine-readable format.
Furthermore, the European Commission has identified the strong need to \enquote{opening up by default all scientific data} and to store and maintain it in the European Open Science Cloud\footnote{\url{http://www.europa.eu/rapid/press-release_IP-16-1408_en}\lastAccessed}. 
The European Union Open Data Portal \footnote{\url{https://www.europeandataportal.eu/en}\lastAccessed} serves as a single point of access to Open Data produced by EU institutions and bodies.
In addition to major data portals, there exists a variety of small data providers.
These data providers either provide data directly as RDF or embed Microformats in their websites.
To automatically find and evaluate small data providers, we extend an existing pipeline to integrate Linked Open Data~\cite{Blume2019}.

\subsection{Open Processing Pipeline}
\label{sec:processing-pipeline}
Information in legal documents is hidden in the text and needs to be extracted to produce legal data.
With this legal data, we can, \eg implement question answering systems and structured text search. 
Furthermore, legal data can be linked to different (external) data sources to provide, \eg background information or geo-locations.

We provide technologies to minimize the effort for tasks that can be applied in a semi-automatic setting.
More specifically, we are interested in the tasks of 
\begin{inparaenum}[(1)]
\item reference extraction, 
\item entity extraction and linking, 
\item keyword and title generation,
\item information retrieval,
\item and visualizing networks.
\end{inparaenum}
The individual components are combined in our open processing pipeline,
whereby the term ``open'' refers to the fact that each component is integrated either as Python package or as external service over the API.
With this approach, the components act as building blocks and can also be used in other projects.

\paragraph{Reference Extraction.}
Reference extraction to legislation and judicial decisions is of great interest~\cite{DeMaat2006}.
A network analysis build on top of citation can reveal decisions with great influence~\cite{Neale2013}.
We extract citations with a hybrid approach that combines rule-based methods with learning-based methods\footnote{\url{https://github.com/openlegaldata/legal-reference-extraction}\lastAccessed}.

\begin{figure}[!bt]
\centering
\includegraphics[page=1,width=.49\textwidth,trim=1cm 0cm 0cm 0cm, clip]{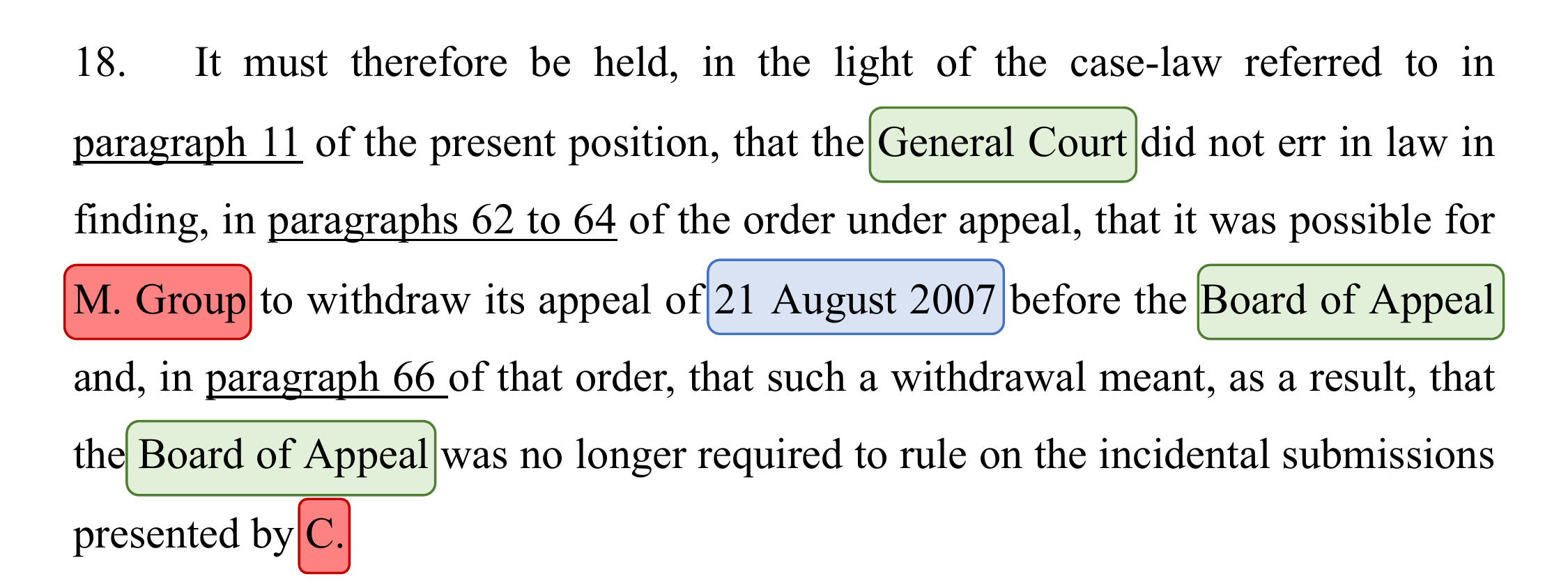}
\caption{\label{fig:legal-ner}Legal NER: Named entities like involved parties (red), organisations (green) or dates (blue) are automatically extracted for text documents like court decisions.}
\end{figure}

\paragraph{Named Entity Recognition \& Entity Linking.}
Extracting named entities such as locations, courts, dates, and times is a well-known information extraction task. 
We implemented this task based on the SpaCy framework\footnote{\url{http://www.spacy.io}\lastAccessed} and trained a German NER model based on the dataset provided by Leitner et al.~\cite{Leitner2019}. %
Extracted entities need to be disambiguated to provide further cross-connections between documents as well as to external data sources.
For example, we link mentions of locations such as cities or states with open geo-information systems like Linked Geo Data\footnote{\url{http://linkedgeodata.org/}\lastAccessed} using the Nominatim service\footnote{\url{https://wiki.openstreetmap.org/wiki/Nominatim}\lastAccessed}.

\paragraph{Keyword and Title Generation.}
Keywords accurately describing the content of documents are of great value for legal information systems. 
To generate keywords, we implement a module that combines rule-based methods with statistical methods.
In Germany, court decision identifiers follow strict rules that allow determining the court and the general domain. 
Thus, we can extract the general domain of court decision by parsing the identifiers. 
Furthermore, referenced laws indicate a predominant legal domain, \eg civil law.
Finally, statistical methods like TF-IDF in combination with thesauri (CF-IDF) allow to generate accurate keywords by analyzing the content (full-text)~\cite{DBLP:conf/kcap/GalkeMSBS17}. 
We apply CF-IDF to generate keywords from legal texts. 
Overall, we obtain keywords by parsing the identifier, analyzing the references to laws, and by analyzing the content of the court decision.
Finally, we combine these keywords to create human-readable titles.

\subsection{Services}
We equip users and developers with the necessary interfaces to efficiently interact with the documents and data. 

\paragraph{Information Retrieval System.}
Finding topically related court decisions is a crucial task for legal professionals.
We implemented a full-text search based on Elasticsearch. 
Furthermore, we developed a text- and citation-based recommender system that assists users in finding relevant information~\cite{Schwarzer2017}.
To facilitate research in this area, the platform provides an open interface such that novel methods can be evaluated with real users in A/B test experiments.

\paragraph{Working Environment.}
\label{sec:annotations}
Typical users access legal information systems with a particular purpose in mind.
While data analysis is an important task, more frequent users are interested in documents regarding a specific topic. 
Finding these legal documents is, however, only the first step in a more complex workflow.
The working environment integrates the hypothesis framework \footnote{\url{http://www.hypothes.is}\lastAccessed}.
Essential pieces of information and extracted entities are highlighted automatically.
Furthermore, users can interact with the documents by highlighting their own key phrases or making notes.

\begin{figure*}[!bt]
\centering
\includegraphics[page=1,width=.99\textwidth,trim=0.75cm 2.5cm 2.5cm 1.0cm, clip]{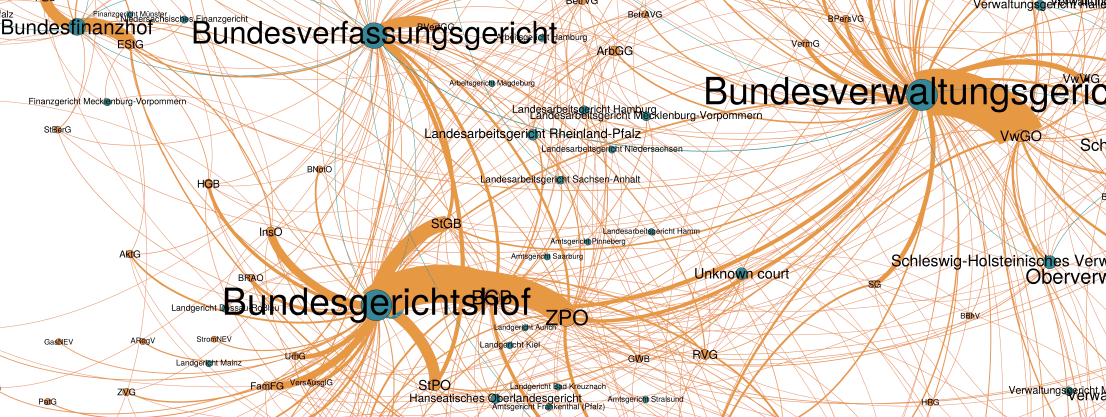}

\caption{\label{fig:citation-network}Excerpt of our citation network of German court decisions\protect\footnotemark. Blue vertices represent courts and orange vertices represent statues books. The size of the vertices indicates the number of court decisions or number of statues available. Blue edges visualize citations from a court decision to another court decision (of the respected court) and orange edges represent citations from court decisions to statue of the respected statue book. The width of the edges indicates the amount of citations.} %
\end{figure*}
\paragraph{Visualization.}
In \cref{fig:citation-network}, we present an excerpt of our citation network generated from German court decisions.
As one can see, federal courts like the \enquote{Bundesverfassungsgericht}, the \enquote{Bundesgerichtshof}, and the \enquote{Bundesverwaltungsgericht} take a central role in the network.
Furthermore, we observe that the \enquote{Bundesverwaltungsgericht} cites the statue book \enquote{Verwaltungsgerichtsordnung} (VwGO) the most.
Our preliminary analysis indicates that many citations to the VwGO are to cite reasons for rejections of revisions.
Moreover, the \enquote{Bundesgerichtshof} cites the statue book \enquote{Bürgerliche Gesetzbuch} (BGB) and the statue book \enquote{Zivilprozessordnung} (ZPO) the most. 
This indicates that most decisions the highest court in Germany are in the civil procedure. 
Citing the ZPO can also indicate a rejection of revision, \eg ZPO § 561.
Additionally, we observe a high amount of citations from the \enquote{Bundesverfassungsgericht} (BVerfG) to the statue book \enquote{Gesetz über das Bundesverfassungsgericht} (BVerfGG), but also to the \enquote{Strafgesetzbuch} (StGB) and the \enquote{Strafprozeßordnung} (StPO).
The BVerfGG contains statues explicitly regulating the BVerfG.
Citations to the StGB and the StPO indicate that the court decision is in the criminal procedure.

Overall, we made some interesting observations in the data by visualizing it, which motivates us to conduct an extensive analysis of our citation network.
However, our analysis has several limitations with regard to the data.
For one, the high amount of court decisions on federal level more likely reflects a more open publishing attitude of federal courts rather than an actual higher workload. 
Thus, it is important to address the publishing policy of courts and raise awareness of the benefits of open access in the justice domain.

\footnotetext{\url{http://openlegaldata.io/assets/pdfs/citation-network.pdf}\lastAccessed}

\section{Conclusion and Outlook}
In this paper, we presented our approach to a single technology stack that allows accessing, processing, and providing legal information. 
We demonstrated that it is feasible to implement a variety of technologies in a single processing pipeline.
We described in detail technologies that are implemented or currently being implemented in our open source project Open Legal Data. 
Furthermore, we published our first dataset of German court decisions\footnote{\url{https://openlegaldata.io/research/2019/02/19/court-decision-dataset.html}\lastAccessed}.
Based on this dataset, legal engineers developed the visual query interface VizLaw and were awarded the first place in the Berlin Legal Tech Hackathon 2019 \footnote{\url{http://www.berlinlegal.tech}\lastAccessed}.
In conclusion, we see the Open Legal Data Platform as an important first step towards openness in the legal domain that will ultimately enable more collaboration among researchers and improve access to justice for the general public.
In this context, we consider the MediaWiki software as role model, that powers all Wikipedias and helped to make encyclopedic knowledge available, and envision to achieve something comparable but for the legal domain.
Making data technically open and accessible is only the beginning. 
In the future, we will also focus on Legal Design in order to make the data and information useful and usable not only for professional users, but also understandable and actionable for lay people.

\begin{acks}

The research presented in this article is funded by the German Federal Ministry of Education and Research (BMBF) through the project Software-Sprint (grant no.~01IS16021). 

\end{acks}

\bibliographystyle{ACM-Reference-Format}
\bibliography{references}

\end{document}